\documentclass[12pt]{article}

\usepackage[dvips]{graphicx}

\usepackage{psfrag}

\def\dspace{\baselineskip = 0.30in}

\def\lapproxeq{\lower .7ex\hbox{$\;\stackrel{\textstyle
<}{\sim}\;$}}
\def\gapproxeq{\lower .7ex\hbox{$\;\stackrel{\textstyle
>}{\sim}\;$}}

\begin{document}

\dspace

\begin{titlepage}
\begin{flushright}
BA-05-102 \\
KIAS-P05054
\end{flushright}
\vskip 2cm
\begin{center}
{\Large\bf
 Flipped SU(5) Predicts $\delta T/T$
} \vskip
1cm {\normalsize\bf $^{(a)}$Bumseok
Kyae\footnote{bkyae@kias.re.kr} and $^{(b)}$Qaisar
Shafi\footnote{shafi@bartol.udel.edu} } \vskip 0.5cm {\it
$^{(a)}$School of Physics, Korea Institute for Advanced Study,
\\207-43, Cheongnyangni-Dong, Dongdaemun-Gu, Seoul 130-722, Korea
\\$^{(b)}$Bartol Research Institute, Department of Physics and Astronomy, \\University of
Delaware, Newark,
DE~~19716,~~USA\\[0.1truecm]}

\end{center}
\vskip 2cm


\begin{abstract}

\noindent We discuss hybrid inflation in supersymmetric flipped
$SU(5)$ model such that the cosmic microwave anisotropy $\delta
T/T$ is essentially proportional to $(M /M_{P})^2$, where $M$
denotes the symmetry breaking scale and $M_{P}$ ($=2.4\times
10^{18}$ GeV) is the reduced Planck mass.  The magnitude of $M$
determined from $\delta T/T$ measurements
%
%
can be consistent with the value inferred from the evolution of
$SU(3)$ and $SU(2)$ gauge couplings. In other words, one could
state that flipped $SU(5)$ predicts (more precisely `postdicts')
$\delta T/T$. The scalar spectral index $n_s = 0.993\pm 0.007$,
the scalar to tensor ratio satisfies $r \lapproxeq 10^{-6}$, while
$dn_s/d{\rm ln}k \lapproxeq 4\times 10^{-4}$.

\end{abstract}
\end{titlepage}

\newpage


In a class of realistic supersymmetric (SUSY) models, inflation is
associated with the breaking of some gauge symmetry $G$, such that
the cosmic microwave anisotropy is essentially proportional to
$(M/M_P)^2$, where $M$ ($\sim M_{GUT}= 2-3\times 10^{16}$ GeV)
denotes the symmetry breaking scale~\cite{hybrid}. The simplest
example of $G$ is provided by $U(1)_{B-L}$, and more complicated
examples based on $SU(5)$~\cite{su5inf} and
$SO(10)$~\cite{so10inf} have also been presented. The Higgs sector
in these grand unified models is typically rather complicated, so
that strictly speaking, the scale $gM$ cannot be identified with
the gauge coupling unification scale (here $g$ denotes the gauge
coupling associated with $G$). For instance, in the $SO(10)$
example inflation is associated with the breaking of $U(1)_{B-L}$
rather than its $SU(5)$ subgroup. In this letter, we hope to
overcome this hurdle by identifying $G$ with $SU(5) \times
U(1)_X$, the so-called flipped $SU(5)$ model~\cite{flip}. This
model is known to possess several advantages over standard grand
unified models such as $SU(5)$ and $SO(10)$ itself that have often
been discussed in the literature. A particularly compelling case
is provided by the ease with which the doublet-triplet (D-T)
splitting can be achieved in models based on $G$. Another
potential advantage is the absence of topological defects
especially monopoles that could create cosmological difficulties.

In this letter, we wish to highlight yet another advantage of
models based on $G$, namely the ease with which a predictive
hybrid inflation scenario~\cite{hybrid} can be realized which is
consistent with D-T splitting and works within a minimal Higgs
framework.
This is in contrast with recent attempts to construct analogous
models based on $SU(5)$~\cite{su5inf} and $SO(10)$~\cite{so10inf},
which turn out to require non-minimal Higgs sectors including
gauge singlet fields.
Although the symmetry breaking scale determined from $\delta T/T$
in the latter case turns out to be comparable to the scale
$M_{GUT}$ determined from the evolutions of the three low energy
gauge couplings, an identification of the two scales is not quite
possible, partly arising from the fact that there are extra Higgs
fields, and also the fact that one has to resort to `shifted'
hybrid inflation~\cite{422} to avoid the monopole problem. These
two issues it appears can be nicely evaded in the flipped model,
so that one could argue that $\delta T/T$ is predicted and turns
out to be in excellent agreement with the
observations~\cite{wmap}. This is the first model of inflation we
are aware of in which this claim can be made. Other testable
predictions include $n_s=0.993\pm 0.007$, $dn_s/d{\rm ln}k
\lapproxeq 4\times 10^{-4}$, and the scalar to tensor ratio
$r\lapproxeq 10^{-6}$. A $U(1)_R$ symmetry plays an essential role
in the construction of this predictive inflationary
scenario~\cite{hybrid,LR}. We will find that its presence implies
a ``double seesaw'' mechanism for realizing suitable masses both
for `right' handed and the light neutrinos. Inflation is followed
by (non-thermal) leptogenesis~\cite{yanagida} with the reheat
temperature consistent with the gravitino
constraint~\cite{gravitino}.


Flipped $SU(5)$ ($=SU(5)\times U(1)_X$) is a maximal subgroup of
$SO(10)$, and contains sixteen chiral superfields per family:
\begin{eqnarray}
{\bf 10}_1= \left(\begin{array}{cc} d^c & Q \\
& \nu^c
\end{array}\right)
 ,~~~ {\bf
\overline{5}}_{-3}= \left(\begin{array}{c} u^c
\\ \overline{L}
\end{array}\right)
  ,~~~ {\bf 1}_{5}= e^c ~.
\end{eqnarray}
Here the subscript refers to the $U(1)_X$ charge in the unit of
$\frac{1}{\sqrt{40}}$.\footnote{We normalize $SU(5)$ and $SO(10)$
generators such that ${\rm Tr}T_{SU(5)}^2=\frac{1}{2}$, and ${\rm
Tr}T_{SO(10)}^2=1$, which is consistent with the $U(1)_X$ charge
normalization.} The MSSM hypercharge is given by a linear
combination of a diagonal $SU(5)$ generator and $U(1)_X$ charge
operator;
\begin{eqnarray} \label{hypercharge}
\frac{1}{2}Y=-\frac{1}{5}Z + \frac{1}{5} X ~,
\end{eqnarray}
where $Z={\rm diag.}(-1/3,-1/3,-1/3,1/2,1/2)$ is the MSSM
hypercharge operator in the standard $SU(5)$ model. Comparison
with standard $SU(5)$ reveals the interchanges,
\begin{eqnarray}
u^c \longleftrightarrow d^c~,~~~ e^c \longleftrightarrow \nu^c ~.
\end{eqnarray}
Since $\nu^c$ has replaced $e^c$ in the ${\bf 10}$-plet, the
latter now belongs to the $SU(5)$ singlet representation.
The MSSM electroweak Higgs doublets reside in two five dimensional
representations as follows:
\begin{eqnarray}
{\bf 5}_{-2}=  \left(\begin{array}{c} \overline{D}^c
\\ H_d
\end{array}\right),~~~ {\bf \overline{5}}_{2} = \left(\begin{array}{c} D^c
\\ H_u
\end{array}\right) ~.
\end{eqnarray}
Comparing to the standard $SU(5)$, $H_u$ and $H_d$ are replaced
each other;
\begin{eqnarray}
H_u \longleftrightarrow H_d ~.
\end{eqnarray}

The breaking of $SU(5)\times U(1)_X$ to the MSSM gauge group is
achieved by providing superlarge VEVs to Higgs of ${\bf 10}$
dimensional representations, namely ${\bf 10}_H$ and ${\bf
\overline{10}}_H$ along the $\nu^c$, $\overline{\nu}^c$
directions. We will provide a very simple superpotential shortly
showing how this is achieved. But first let us briefly recall how
the D-T splitting problem is elegantly solved in this framework.
Consider the superpotential couplings,
\begin{eqnarray}
{\bf 10}_H {\bf 10}_H {\bf 5}_{-2}~~~ {\rm and}~~~ {\bf
\overline{10}}_H {\bf \overline{10}}_H {\bf \overline{5}}_2 ~.
\end{eqnarray}
With superlarge VEVs of ${\bf 10}_H$ and ${\bf\overline{10}}_H$
along the $\nu^c_H$, $\overline{\nu}^c_H$ directions respectively,
we see that $\overline{D}^c$ and $D^c$ in the ${\bf 5}_{-2}$ and
${\bf\overline{5}}_2$ pair up to be superheavy with their
corresponding partners, $d^c_H$ and $\overline{d}^c_H$ from ${\bf
10}_H$ and ${\bf\overline{10}}_H$. [$Q_H$ and $\overline{Q}_H$
contained in ${\bf 10}_H$ and ${\bf\overline{10}}_H$ are, on the
other hand, absorbed by the gauge sector, when flipped $SU(5)$ is
broken to the MSSM gauge group.] Thus, the electroweak Higgs
doublets remain unpaired as desired, as long as a bare mass term
${\bf 5}_{-2}{\bf\overline{5}}_2$ with coefficient of order
$M_{GUT}$, permitted by the gauge symmetry can be avoided.
Indeed, if this coefficient can be of order $M_W$ rather than
$M_{GUT}$, we would achieve two worthwhile goals. Namely, the MSSM
$\mu$ problem would be resolved and dimension five nucleon decay
would be essentially eliminated. The need for some additional
symmetry is also mandated by our desire to implement a predictive
inflationary scenario along the lines discussed in earlier
papers~\cite{hybrid,su5inf,so10inf,422,LR}. We have found that a
$U(1)_R$ symmetry is particularly potent in constraining the
inflationary superpotential and will therefore exploit it here.

Disregarding the pure right handed neutrino sector for the moment,
the superpotential responsible for breaking the $SU(5)\times
U(1)_X$ gauge symmetry, resolving the D-T splitting problem, and
generating Dirac mass terms for the charged fermions and neutrinos
is as follows:
\vskip 0.6 cm
\begin{eqnarray} \label{superpot}
&&~~~~~~~~~~~W=\kappa S\left[{\bf
10}_H{\bf\overline{10}}_H-M^2\right]
\nonumber \\
&&~~~~~~~+\lambda_1{\bf 10}_H{\bf 10}_H{\bf
5}_h+\lambda_2\overline{{\bf 10}}_H\overline{{\bf
10}}_H\overline{{\bf 5}}_h
\\
&&+y_{ij}^{(d)}{\bf 10}_i{\bf 10}_j{\bf 5}_h +y_{ij}^{(u,\nu)}{\bf
10}_i\overline{{\bf 5}}_j\overline{{\bf 5}}_h + y_{ij}^{(e)}{\bf
1}_i\overline{{\bf 5}}_j{\bf 5}_h \nonumber  ~.
\end{eqnarray}
The quantum numbers of the superfields appearing in
Eq.~(\ref{superpot}) are listed in Table I.
\begin{center}
\begin{tabular}{|c||cc|cc|cc|ccc|} \hline
& ~$S$ & $\Sigma$ & ${\bf 10}_H$ & ${\bf \overline{10}}_H$ & ${\bf
5}_h$ & ${\bf\overline{5}}_h$ & ${\bf 10}_i$ &
${\bf\overline{5}}_i$ & ${\bf 1}_i$
\\ \hline\hline
$X$ & $0$ & $0$ & $1$ & $-1$ & $-2$ & $2$ & $1$ & $-3$ & $5$
\\
$R$ & $1$ & $2$ & $0$ & $0$ & $1$ & $1$ & $0$ & $0$ & $0$
\\
$Z_2$ & $+$ & $+$ & $+$ & $+$ & $+$ & $+$ & $-$ & $-$ & $-$
\\ \hline
\end{tabular}
\vskip 0.4cm {\bf Table I~}
\end{center}
The $U(1)_R$ symmetry eliminates terms such as $S^2$ and $S^3$
from the superpotential, which yields a predictive inflationary
scenario~\cite{hybrid}. Higher dimensional baryon number violating
operators such as ${\bf 10}_i{\bf 10}_j{\bf
10}_k{\bf\overline{5}}_l\langle S\rangle/M_P^2$, ${\bf
10}_i{\bf\overline{5}}_j{\bf \overline{5}}_k{\bf 1}_l\langle
S\rangle/M_P^2$, etc. are heavily suppressed as a consequence of
$U(1)_R$. Thus, we expect proton decay to proceed via dimension
six operators mediated by the superheavy gauge bosons. The
dominant decay mode is $p\rightarrow e^+/\mu^+,\pi^0$ and the
estimated lifetime is of order $10^{36}$
yrs~\cite{flipping,pdecay}.

We note that the `matter' superfields  (and ${\bf 10}_H$) are
neutral under $U(1)_R$, so that the $Z_2$ subgroup of the latter
does not play the role of `matter' parity. An additional $Z_2$
symmetry (`matter' parity) has been introduced to avoid
undesirable couplings such as ${\bf 10}_H{\bf 10}_i{\bf 5}_h$,
${\bf 10}_H{\bf \overline{5}}_i{\bf\overline{5}}_h$, ${\bf
10}_H{\bf 10}_i{\bf 10}_j{\bf\overline{5}}_kS$, ${\bf
10}_H{\bf\overline{5}}_i{\bf \overline{5}}_j{\bf 1}_kS$,
etc.\footnote{${\bf 10}_H{\bf 10}_i{\bf 5}_h$ and ${\bf 10}_H{\bf
\overline{5}}_i{\bf\overline{5}}_h$ ($\supset\langle
\nu^c_H\rangle d^c_i\overline{D}^c$, $\langle \nu^c_H\rangle
L_iH_u$) induce superheavy mass terms of $d_i^c$, $L_i$, and
$H_u$. ${\bf 10}_H{\bf 10}_i{\bf 10}_j{\bf\overline{5}}_kS$ and
${\bf10}_H{\bf\overline{5}}_i{\bf \overline{5}}_j{\bf 1}_kS$
($\supset\langle\nu^c_H\rangle Q_iL_jd^c_k\langle S\rangle/M_P^2$,
$\langle\nu^c_H\rangle d^c_id^c_ju^c_k\langle S\rangle/M_P^2$,
$\langle\nu^c_H\rangle L_iL_je^c_k\langle S\rangle/M_P^2$) lead to
proton as well as LSP decays.}
%
%
This $Z_2$ ensures that the LSP is absolutely stable and
consequently a desirable candidate for CDM.

The superpotential in Eq.~(\ref{superpot}) and the ``D-term''
potential, in the global SUSY limit, possesses a ground state in
which the scalar components (labelled by the same notation as the
corresponding superfield) acquire the following VEVs:
\begin{eqnarray}
|\langle {\bf 10}_H\rangle| = |\langle {\bf
\overline{10}}_H\rangle| = M ~,~~~{\rm and}~~~\langle S\rangle
=0~.
\end{eqnarray}
Thus, the gauge symmetry $SU(5)\times U(1)_X$ is broken at the
scale $M$, while SUSY remains unbroken. In a $N=1$ supergravity
framework, after including the soft SUSY breaking terms, the $S$
field acquires a VEV proportional to the gravitino mass
$m_{3/2}$~\cite{LR}.  As explained above, unwanted triplets in
${\bf 10}$-plet and ${\bf 5}$-plet Higgs become superheavy by the
$\lambda$ couplings in Eq.~(\ref{superpot}). From the
$y^{(d)}_{ij}$ and $y^{(e)}_{ij}$ terms in Eq.~(\ref{superpot}),
d-type quarks and charged leptons acquire masses after electroweak
symmetry broken. $y^{(u,\nu)}_{ij}$ term provides u-type quarks
and neutrino with (Dirac) masses. Since $u^c_i$ and $L_i$ are
contained in a single multiplet ${\bf \overline{5}}_i$, the mass
matrices for the u-type quarks and Dirac neutrinos are related in
flipped $SU(5)$:
\begin{eqnarray}
M^u_{ij}=M^{\nu}_{ji}=y^{(u,\nu)}_{ij}\langle H_u\rangle ~.
\end{eqnarray}

We point out here that $U(1)_R$ forbids the bare mass term ${\bf
5}_{h}{\bf\overline{5}}_h$, so that the electroweak Higgs doublets
do not acquire superheavy masses. To resolve the MSSM $\mu$
problem we invoke, following~\cite{mu}, the following term in the
K${\rm\ddot{a}}$hler potential:
\begin{eqnarray} \label{mu}
K\supset y_{\mu}\frac{\Sigma^\dagger}{M_P} {\bf 5}_h\overline{{\bf
5}}_h +{\rm h.c.} ~.
\end{eqnarray}
Intermediate scale SUSY breaking triggered by the hidden sector
superfield $\Sigma$ via $\langle F_\Sigma\rangle\sim m_{3/2}M_P$
yields the MSSM $\mu$ term, $\mu\equiv
y_{\mu}\frac{F_\Sigma^*}{M_P}\sim m_{3/2}$. The quantum numbers of
$\Sigma$ are listed in Table I.

To realize the simplest inflationary scenario, the scalars must be
displaced from their present day minima. Thus, for $\langle
S\rangle >>M$, the scalars ${\bf 10}_H$ and ${\bf
\overline{10}}_H\rightarrow 0$, so that the gauge symmetry is
restored but SUSY is broken. This generates a tree level scalar
potential $V_{\rm tree}=\kappa^2 M^4$, which will drive inflation.
In practice, in addition to the supergravity corrections and the
soft SUSY breaking terms, we also must include one loop radiative
corrections arising from the fact that SUSY is broken by $\langle
S\rangle\neq 0$ during inflation. This causes a split between the
masses of the scalar and fermionic components in ${\bf 10}_H$,
${\bf \overline{10}}_H$. For completeness, following
Refs.~\cite{ns,ns2}, we provide here the inflationary potential
that is employed to compute the CMB anisotropy $\delta T/T$, the
scalar spectral index $n_s$, and the tensor to scalar ratio $r$:
\begin{eqnarray} \label{infpot}
&&V =V_{\rm tree}\times \bigg[1+ \frac{\kappa^2{\cal
N}}{32\pi^2}\bigg(2{\rm ln}\frac{\kappa^2|S|^2}{\Lambda^2}
+(1+z)^2{\rm ln}(1+z^{-1})
\nonumber \\
&&~~~~ +(1-z^{-1})^2{\rm
ln}(1-z^{-1})\bigg)+\frac{|S|^4}{2M_P^4}\bigg] + a m_{3/2}\kappa
M^2|S|^2  ~,
\end{eqnarray}
where $z\equiv |S^2|/M^2$, ${\cal N}$ ($=10$) denotes the
dimensionality of ${\bf 10}_H$, ${\bf \overline{10}}_H$, and
$\Lambda$ is a renormalization mass. We have employed a minimal
K${\rm\ddot{a}}$hler potential and $a=2|2-A|~ {\rm cos}[{\rm
arg}S+{\rm arg}(2-A)]$, where $A$ denotes the ``A-parmater''
associated with the soft terms. Note that during the last 60 or so
e-foldings the value of the $S$ field is well below $M_P$,
especially for $\kappa \lapproxeq 10^{-2}$, which means that the
supergravity correction proportional to $|S|^4/ M_P^2$ is
adequately suppressed (see Fig. 1).  In addition, the soft term
also can be safely ignored for $\kappa \gapproxeq 10^{-3}$.

Neglecting the supergravity correction and the soft term in
Eq.~(\ref{infpot}), $\delta T/T$ is given by~\cite{hybrid}
\begin{eqnarray} \label{appx}
\frac{\delta T}{T} \approx \sqrt{\frac{N_l}{45 {\cal N}}}\times
\left(\frac{M}{M_P}\right)^2 ~,
\end{eqnarray}
where we took $z<<M$. $N_l$ indicates the number of e-foldings
(=50--60) and ${\cal N}=10$ as mentioned above. Substituting
$\delta T/T \sim 6 \times 10^{-6}$ (corresponding to the comoving
wave number $k_0=0.002$ $\rm{Mpc^{-1}}$), one estimates $M$ to be
of order $10^{16}$ GeV. Alternatively, one could insert in
Eq.~(\ref{appx}) the magnitude of $M$ determined from the symmetry
breaking of flipped $SU(5)$ and thereby `predict' $\delta T/T$.
There is good agreement one finds with the observations.

To make a more precise comparison between $M$ determined from
$\delta T/T$ and its value determined from the evolution of the
$SU(3)$ and $SU(2)$ gauge couplings, we should include the
supergravity corrections as well as the corrections coming from
the soft terms. The results are exhibited in Fig. 1.
%

The unification of $SU(3)$ and $SU(2)$ gauge couplings occurs at
the scale $g_5 M$, where $g_5$ ($\approx 0.7$) denotes the $SU(5)$
gauge coupling.\footnote{In flipped $SU(5)$, $\sqrt{5/3}g_Y$ is
not unified with $g_5$ at the scale $g_5M$. The $SU(5)$ and
$U(1)_X$ gauge couplings continue to evolve above the scale
$g_5M$, and eventually are unified at a higher scale $M_U$. The
consistency condition $g_5M\leq M_U$ implies $g_{X}\leq g_5$ at
the scale $g_5M$, from which we obtain the generic upper bound,
$g_5M\leq M_G$~\cite{flipping}. Here $M_G$ ($\approx 2.03\times
10^{16}$ GeV) denotes the standard $SU(5)$ unification scale.}
In Fig. 1 we plot $M$ versus $\kappa$, which shows that the
coupling unification scale $g_5 M$ lies in the range $3.8\times
10^{15} - 1.4 \times 10^{16}$ GeV for $10^{-5}\lapproxeq \kappa
\lapproxeq 10^{-2}$.
$g_5M$ determined from the two loop RG evolutions of the MSSM
gauge couplings with the initial values $g_3^2/4\pi(M_Z)=0.1187\pm
0.002$ and ${\rm sin}^2\theta_W^{\overline{MS}}(M_Z)=0.23120\pm
0.00015$~\cite{PDG} turns out to be $6.1\times 10^{15}~{\rm GeV
}\lapproxeq g_5M\lapproxeq 1.02\times 10^{16}~{\rm GeV}$ (or
$8.7\times 10^{15}~{\rm GeV }\lapproxeq M\lapproxeq 1.46 \times
10^{16}~{\rm GeV}$)~\cite{flipping}.
Values of $\kappa$ of order $10^{-2}-10^{-3}$ and $10^{-5}$ are in
good agreement with this.
%
%
To quantify this somewhat differently, in Fig. 2 we plot the
predicted curvature perturbation (for $k_0 = 0.002 ~{\rm
Mpc^{-1}}$)
\begin{eqnarray} \label{R}
{\cal R} = \frac{1}{2\sqrt{3}\pi M_P^3}~\frac{V^{3/2}}{|V'|}
\end{eqnarray}
as a function of $\kappa$, for varying values of the symmetry
breaking scale $M$. Good agreement with observations is possible
for $\kappa$ around $10^{-5}$, or $\kappa$ in the vicinity of
$10^{-2}$~\cite{ns,ns2}. Note that for $\kappa$ larger than
$2\times 10^{-2}$, the scalar spectral index exceeds unity due to
supergravity corrections. Thus, values of $M$ close to $10^{16}$
GeV are preferred if $\kappa$ is in the vicinity of $10^{-2}$.
Note that for $\kappa\sim 10^{-2}-10^{-3}$, the supergravity
corrections and soft terms are safely neglected, in which case,
from Eq.~(\ref{appx}), $\delta T/T$ is `predicted.'

From Fig. 2 we note that the region $\kappa\sim (2-3)\times
10^{-5}$ also leads to ${\cal R}$ in good agreement with the
observations. While this cannot be claimed as a `prediction,' it
can be experimentally distinguished from the region $\kappa\sim
10^{-2}-10^{-3}$ by a precise measurement of the scalar spectral
index and other quantities which we now discuss. (Note that for
$\kappa$ less than or of order few $\times 10^{-6}$, ${\cal R}$
rapidly falls below the observed value for all plausible values of
the symmetry breaking scale $M$).

Fig. 3 displays the dependence of the scalar spectral index $n_s$
on $\kappa$. With $\kappa \lapproxeq 10^{-2}$ required by the
constraint $T_r \lapproxeq 10^9$ GeV, we find that $n_s = 0.993
\pm 0.007$. Measurement of $n_s$ to better than a percent is
eagerly awaited. Fig. 4 shows that $dn_s/ d{\rm ln}k\lapproxeq
4\times 10^{-4}$. The tensor to scalar ratio $r\lapproxeq 10^{-6}$
(Fig. 5).

The end of inflation is reached when the scalar field $S$ leaves
the `slow roll' regime and rapidly approaches its true minimum
near the origin. This is also the signal for the fields ${\bf
10}_H$, ${\bf\overline{10}}_H$ to leave their positions at the
origin and proceed to their minima at $M$.  Since the breaking of
$SU(5)\times U(1)_X$ does not produce monopoles, there are no
cosmological problems to worry about, as stated earlier. The
scalar fields perform damped oscillations about their minima and
eventually decay, leading to reheat temperature $T_r \lapproxeq
10^9$ GeV, in order to the gravitino problem is avoided.\footnote{
Recently it was argued that the reheat temperature $T_r\lapproxeq
10^{6-7}$ GeV for $m_{3/2}\sim 100$ GeV, unless the hadronic
decays of the gravitino are suppressed~\cite{moroi}.
The low temperature leptogenesis scenario ($T_r \sim 10^2$ GeV)
could work in this class of inflationary models~\cite{ns2}, if two
right handed neutrinos with masses of $10^4$ GeV or so are nearly
degenerate.
Note that the heaviest right handed neutrino can still have a much
larger mass than the inflaton (say $\sim 10^{14}$ GeV; thus it
cannot be produced by the inflaton perturbative decay).
The baryon asymmetry is then of order $(T_r/ m_\chi)\times
\epsilon$, where $m_\chi$ denotes the inflaton mass and $\epsilon$
the lepton asymmetry per neutrino decay. $\kappa \sim 10^{-5}$
(thus $m_\chi\sim 10^{11}$ GeV) and $\epsilon\sim O(1)$ give the
desired baryon asymmetry ($\sim 10^{-10}$).
Actually, $\epsilon$ can be as large as 1/2, provided the neutrino
mass splittings are comparable to their decay
widths~\cite{pilaftsis}.
%
%
The constraint $T_r\lapproxeq 10^9$ GeV remains intact if either
the hadronic decay ratio of the gravitino is small enough
($\lapproxeq 10^{-3}$) and $m_{3/2}\gapproxeq 3$ TeV, or if the
gravitino happens to be the LSP.
The gravitino, in principle, could have many decay channels into
hidden sector fields, which would be helpful for lowering its
hadronic decay ratio.  Moreover, if the gravitino is the LSP and
$T_r\sim 10^{10}$ GeV, the gravitino can be the dark matter in the
universe~\cite{bolz}.}
See Fig. 5 for the dependence of $T_r$ on $\kappa$. It shows that
$\kappa \lapproxeq 10^{-2}$ for $T_r \lapproxeq 10^9$ GeV.

We expect the decay to proceed via the production of right handed
neutrinos and sneutrinos arising from the quartic (dimension five)
superpotential couplings ${\bf 10}_i {\bf 10}_j
{\bf\overline{10}}_H {\bf\overline{10}}_H$, in combination with
the coupling $S{\bf 10}_H {\bf\overline{10}}_H$. The former
coupling is permitted by the $SU(5)\times U(1)_X$ symmetry and
would normally give rise to large ($\lapproxeq 10^{14}$ GeV) right
handed neutrino masses.
%
Assuming hierarchical right handed neutrino masses,
%
such couplings typically give rise to a reheat temperature $T_r$
of order $(1/10-1/100) M_N$, where $M_N$ denotes the mass of the
heaviest right handed neutrino that can be produced by the
decaying inflaton~\cite{lepto-inf}. Furthermore, the decaying
right handed neutrinos can provide a nice explanation of the
observed baryon asymmetry via leptogenesis~\cite{yanagida}.
%
Following Ref.~\cite{ns2}, one can derive the $\kappa$
dependence of $T_r$ shown in Fig. 5.
%
The quartic coupling above is, however, inconsistent with the
$U(1)_R$ symmetry, and a somewhat more elaborate scenario based on
the ``double seesaw''~\cite{dblss} is required to implement both
the usual seesaw mechanism and the desired reheat scenario. The
details of one such extension are as follows.

With additional $SU(5)\times U(1)_X$ singlets superfields $\Phi$,
$\Phi^{'}$, and $\Psi$ and a hidden (anomalous) gauge symmetry
$U(1)_H$, let us consider the superpotential:
\begin{eqnarray}
W_{\nu^c}= \frac{\rho_{i}}{M_P}\Phi\Psi{\bf 10}_i\overline{{\bf
10}}_H + \frac{\rho}{M_P}\Phi\Phi'\Psi\Psi ~,
\end{eqnarray}
where $\rho_i$, $\rho$ denote dimensionless coupling constants.
The quantum numbers of $\Phi$, $\Phi^{'}$, and $\Psi$ are shown in
Table II.
%
%
\begin{center}
\begin{tabular}{|c||ccc|} \hline
 & $\Phi$ & $\Phi'$ & $\Psi$
 \\ \hline\hline
$X$ & $0$ & $0$ & $0$   \\
$R$ & $0$ & $-1$ & $1$
\\
$Z_2$ & $+$ & $+$ & $-$
\\ \hline
$H$ & $1$ & $1$ & $-1$
\\
\hline
\end{tabular}
\vskip 0.4cm {\bf Table II~}
\end{center}
Note that the superfields appearing in Table I are neutral under
$U(1)_H$.
From the ``D-term'' scalar potential associated with $U(1)_H$,
\begin{eqnarray}
V_D=\frac{g_H^2}{2}\bigg||\Phi|^2+|\Phi'|^2-|\Psi|^2-\xi\bigg|^2
~,
\end{eqnarray}
the scalar components of $\Phi$ and $\Phi'$ develop non-zero VEVs;
$\sqrt{|\Phi|^2+|\Phi'|^2}=\sqrt{\xi}\sim 10^{17}$ GeV, while
$\langle \Psi\rangle$ vanishes by including the soft terms in the
potential.  Here the parameter $\xi$ comes from the
``Fayet-Iliopoulos D-term''~\cite{FI}. Since $\xi >>
\kappa^2M^4/M_P^2$, $U(1)_H$ is broken even during inflation.
Thus, cosmic strings associated with $U(1)_H$ breaking would be
inflated away.

By including soft SUSY breaking terms and supergravity effects
from the higher order K${\rm\ddot{a}}$hler potential term,
\begin{eqnarray} \label{highK}
K\supset \frac{h}{M_P^2}\Phi\Phi^\dagger\Phi'\Phi^{'\dagger} ~,
\end{eqnarray}
where $h$ is real and $-1\lapproxeq h < 0$, $\langle \Phi\rangle$
and $\langle \Phi'\rangle$ could be determined such that $|\langle
\Phi\rangle| = |\langle \Phi'\rangle|$. We have assumed here that
terms proportional to $(\Phi\Phi^\dagger)^2$ and
$(\Phi'\Phi^{'\dagger})^2$ in the K${\rm\ddot{a}}$hler potential
are suppressed relative to the one given in Eq.~(\ref{highK}). It
turns out that $|\langle \Phi\rangle| = |\langle \Phi'\rangle|$
also holds during inflation. The VEV of the lighter mass
eigenstate [$=(\Phi-\Phi')/\sqrt{2}$] vanishes both during and
after inflation, while the superheavy mass eigenstate
[$=(\Phi+\Phi')/\sqrt{2}$] develops a VEV of order
$\sqrt{\xi}$.\footnote{Alternatively, $\langle \Phi\rangle$ and
$\langle \Phi'\rangle$ could be determined by introduction of
$\overline{\Phi}$ with proper quantum numbers assigned and the
nonrenormalizable terms in the superpotential, $W\supset
S(\kappa'\Phi\overline{\Phi}-\rho'(\Phi\overline{\Phi})^2/M_P^2)$.
By including soft terms, it turns out $\langle \Phi\rangle =
\langle \overline{\Phi}\rangle= \sqrt{\kappa'M_P^2/\rho'}$ at a
local minimum~\cite{so10inf}. From the ``D-term'' potential, we
have $|\langle \Phi'\rangle|^2=\xi$. We should assmue
$\kappa'/\rho'<<1$ to keep intact the inflationary scenario
discussed so far. In this case also the VEV of the lighter mass
eigenstate vanishes both during and after inflation.}
%
%
%

With $\langle \Phi\rangle\sim \langle \Phi'\rangle\sim 10^{17}$
GeV, $\Psi$ has a superheavy Majorana mass of order
$\rho\langle\Phi\Phi'\rangle/M_P \sim 10^{16}$ GeV, while $\Psi$
and $\nu_i^c$ obtain (psuedo-) Dirac masses of order
$\rho_i\langle \Phi\rangle\langle \overline{\nu}^c_H\rangle/M_P
\lapproxeq 10^{15}$ GeV. Hence, the ``seesaw masses'' ($\sim
[\rho_i^2\langle\Phi\rangle/\rho\langle\Phi'\rangle]\times
[\langle\overline{\nu}^c_H\rangle^2/ M_P]$) of the lighter mass
eigenstates, which are indeed the ``physical'' right handed
neutrinos, should be of order $10^{14}$ GeV or smaller, as
desired.


In summary, it is tempting to think that inflation is somehow
linked to grand unification, especially since the scale associated
with the vacuum energy that drives inflation should be less than
or of order $10^{16}$ GeV. Models in which $\delta T/T$ is
proportional to $(M/M_P)^2$, with $M$ comparable to $M_{GUT}$ are
especially interesting in this regard~\cite{hybrid}.
Supersymmetric flipped $SU(5)$ provides a particularly compelling
example in which the magnitude of $\delta T/T$ can be `predicted'
by exploiting the gauge coupling unification scale that has been
known for several years. Precise measurement of the scalar
spectral index will provide an important test of this class of
models.

\vskip 0.7cm
\noindent {\bf Acknowledgments}

\noindent
We thank Nefer Senoguz for helpful discussions and for providing
us with the figures. Q.S. is partially supported by the DOE under
contract No. DE-FG02-91ER40626. Q.S. acknowledges the hospitality
of Professors Eung Jin Chun and Chung Wook Kim at KIAS where this
project was initiated.




\begin{figure}[htb]
\includegraphics[angle=0, width=14cm]{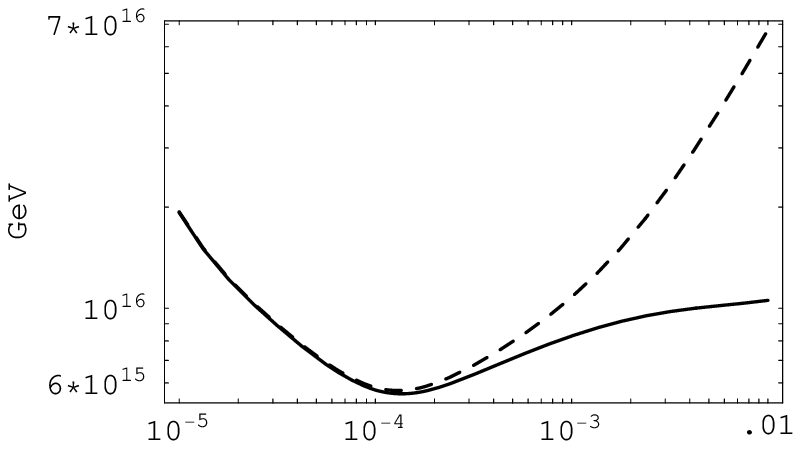}
\vspace{-1.0cm}
\begin{center}
{\large \qquad $\kappa$}
\end{center}
 \vspace{-0.7cm}
\caption{\sf The value of the symmetry breaking scale $M$ (solid)
and the magnitude of the inflaton $|S|$ (dashed) vs. $\kappa$. We
take $m_{3/2}=10^3$ GeV and $a>0$.} \label{fig:21}
\end{figure}

\begin{figure}[htb]
\psfrag{R}{\scriptsize{$\mathcal{R}$}}
\includegraphics[angle=0, width=14cm]{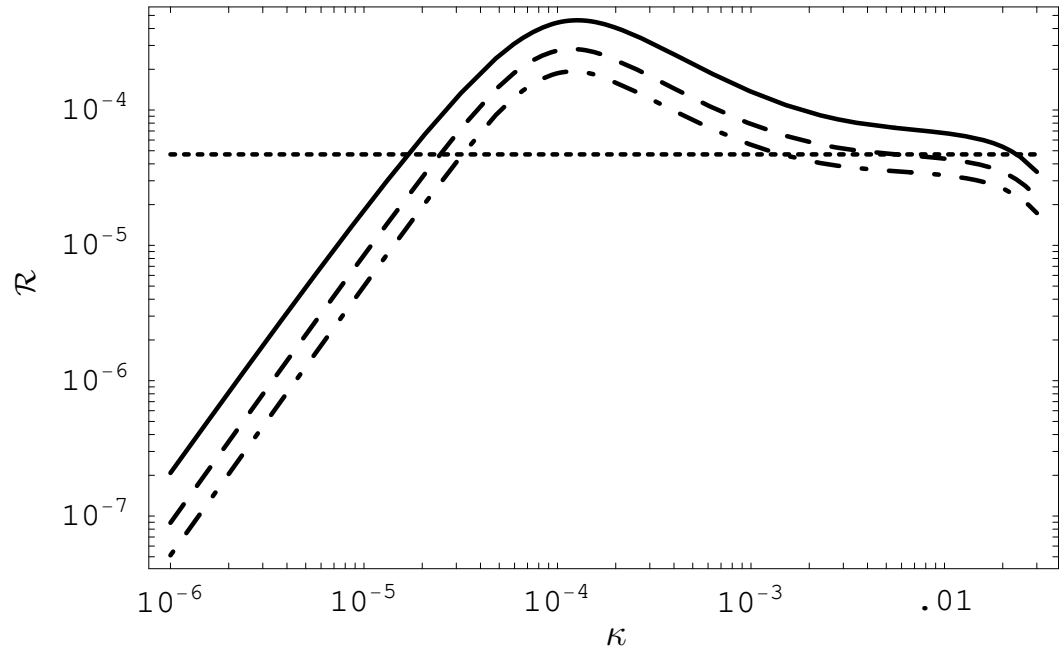}
\vspace{-.8cm}
\begin{center}
{\large \qquad $\kappa$}
\end{center}
\vspace{-0.5cm} \caption{\sf $\mathcal{R}$ vs. $\kappa$.
$M=1.24\times10^{16}$ GeV (solid), $M=1\times10^{16}$ GeV
(dashed), and $M=8.7\times10^{15}$ GeV (dot-dashed). The dotted
horizontal line corresponds to
 $\mathcal{R}=(4.7\pm 0.3)\times10^{-5}$. We take $m_{3/2}=10^3$ GeV and $a>0$.}
\label{fig:21}
\end{figure}

\begin{figure}[htb]
\includegraphics[angle=0, width=14cm]{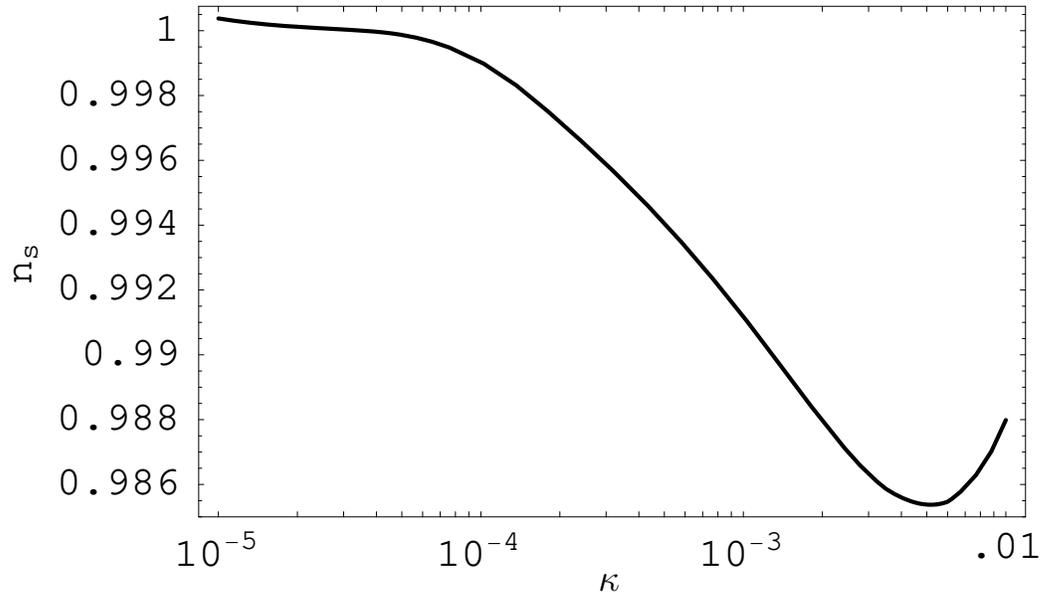}
\vspace{-1.2cm}
\begin{center}
{\large \qquad $\kappa$}
\end{center}
 \vspace{-.5cm}
\caption{\sf The spectral index $n_s$ vs. $\kappa$.}
\label{fig:22}
\end{figure}

\begin{figure}[htb]
\includegraphics[angle=0, width=14cm]{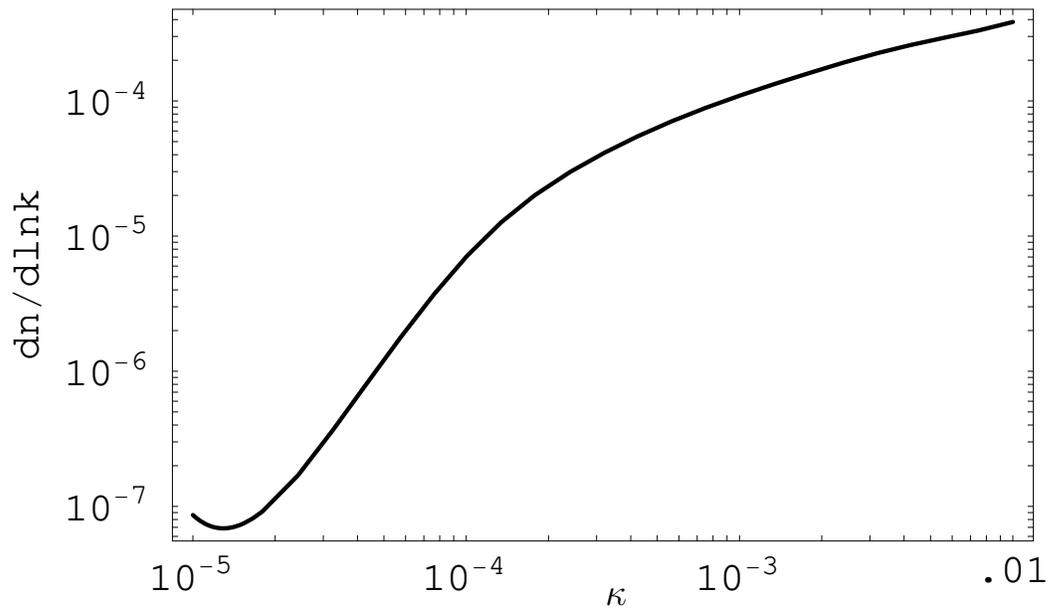}
\vspace{-1.2cm}
\begin{center}
{\large \qquad $\kappa$}
\end{center}
 \vspace{-.5cm}
\caption{\sf ${\rm d}n_s/{\rm d}\ln k$ vs. $\kappa$.}
\label{fig:22}
\end{figure}

\begin{figure}[htb]
\includegraphics[angle=0, width=14cm]{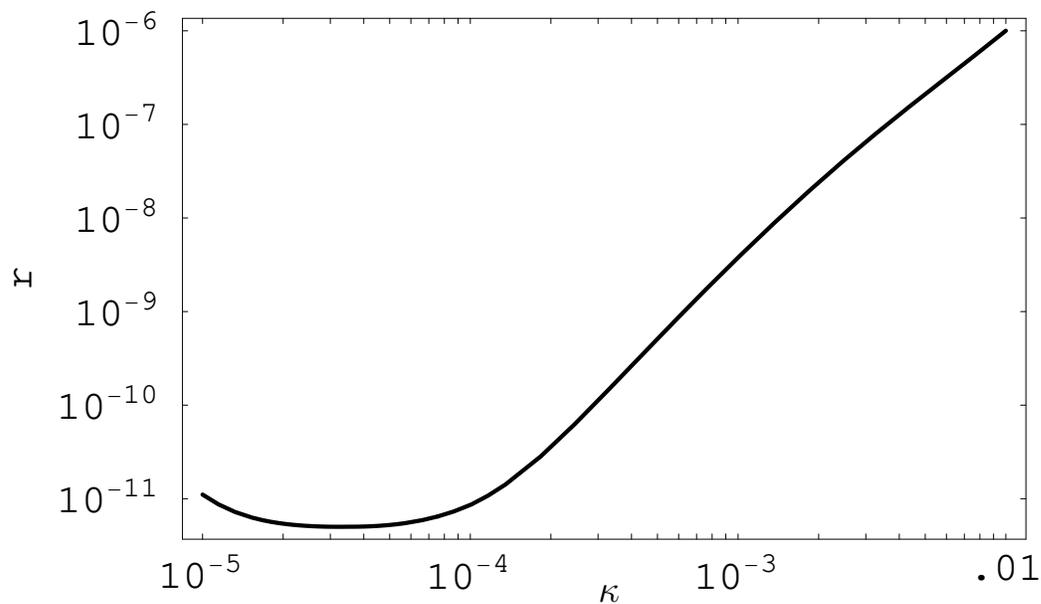}
\vspace{-1.2cm}
\begin{center}
{\large \qquad $\kappa$}
\end{center}
 \vspace{-.5cm}
\caption{\sf The tensor to scalar ratio $r$ vs. $\kappa$.}
\label{fig:22}
\end{figure}

\begin{figure}[htb]
\includegraphics[angle=0, width=14cm]{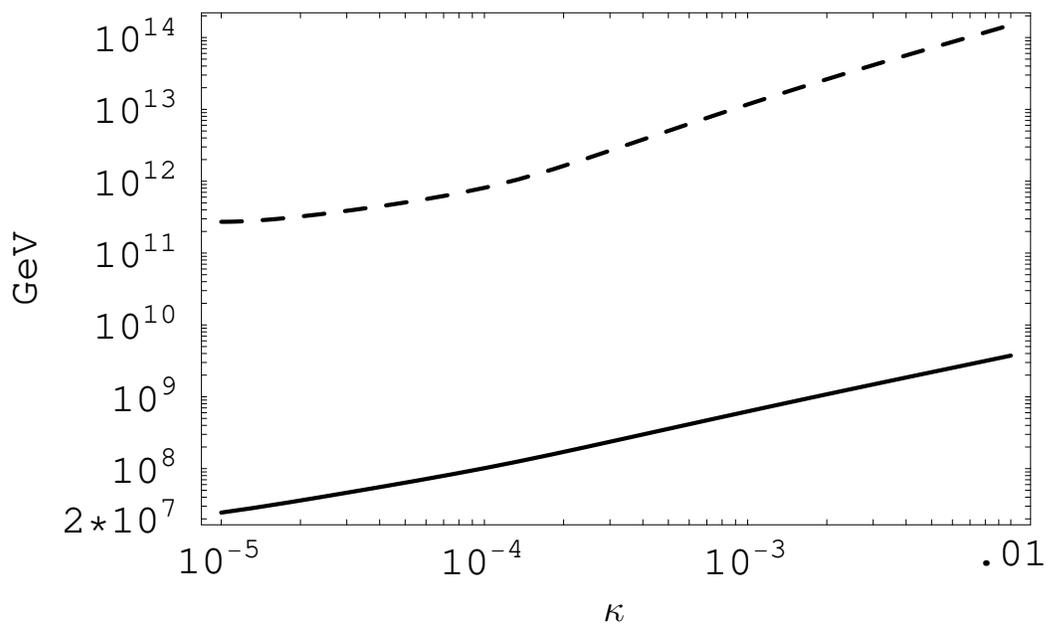}
\vspace{-0.9cm}
\begin{center}
{\large \qquad $\kappa$}
\end{center}
 \vspace{-0.8cm}
\caption{\sf The lower bound on the reheat temperature $T_r$
(solid) and the inflaton mass $m_{\chi}$ (dashed) vs. $\kappa$.}
\label{minf}
\end{figure}


\end{document}